\documentclass[twocolumn,showpacs,preprintnumbers,amsmath,amssymb,apl]{revtex4}
\usepackage{graphicx}
\usepackage{dcolumn}
\usepackage{bm}
\usepackage{color}
\begin{document}
\title{Non-equilibrium Green's function based single-band tight-binding model for Fe-MgO-Fe magnetic tunnel junction devices}
\author{Tehseen Raza}
\affiliation{School of Electrical and Computer Engineering and NSF Network for Computational Nanotechnology, Purdue University, West Lafayette, Indiana 47907}

\author{Hassan Raza}
\affiliation{School of Electrical and Computer Engineering, Cornell University, Ithaca NY 14853}

\begin{abstract}
Motivated by observation of very high tunnel magnetoresistance (TMR) in Fe-MgO-Fe magnetic tunnel junction devices, we propose a theoretical model for these devices based on a single-band tight-binding approximation. An effort is made to capture the band dispersions over the two dimensional transverse Brillouin zone. In the transport direction, spin dependent Hamiltonian is prescribed for $\Delta_1$ and $\Delta_5$ bands. Non-equilibrium Green's function formalism is then used to calculate transport. Features like voltage dependence of TMR are captured quantitatively within this simple model and the trends match well with the ones predicted by \textit{ab-initio} methods and experiments.
\end{abstract}


\maketitle
Magnetic tunnel junction (MTJ) devices have emerged as one of the candidates for random access memory applications. The prediction of high tunnel magnetoresistance (TMR) for crystalline MgO barrier \cite{Butler01, Mathon01} was followed by observations of upto 200$\%$ TMR ratios at room temperature in Fe-MgO-Fe and CoFe-MgO-CoFe MTJ devices \cite{Yuasa04,Parkin04}. Since then, there has been an increased effort to integrate them into practical devices. Although, \textit{ab-initio} \cite{Butler01, Derek07, Heiliger05, Heiliger06} and empirical tight binding \cite{Mathon01} studies have been reported, their computational complexity limit their use for rapid device prototyping.

\begin{figure*}
\vspace{2.2in}
\hskip -3.0in\includegraphics{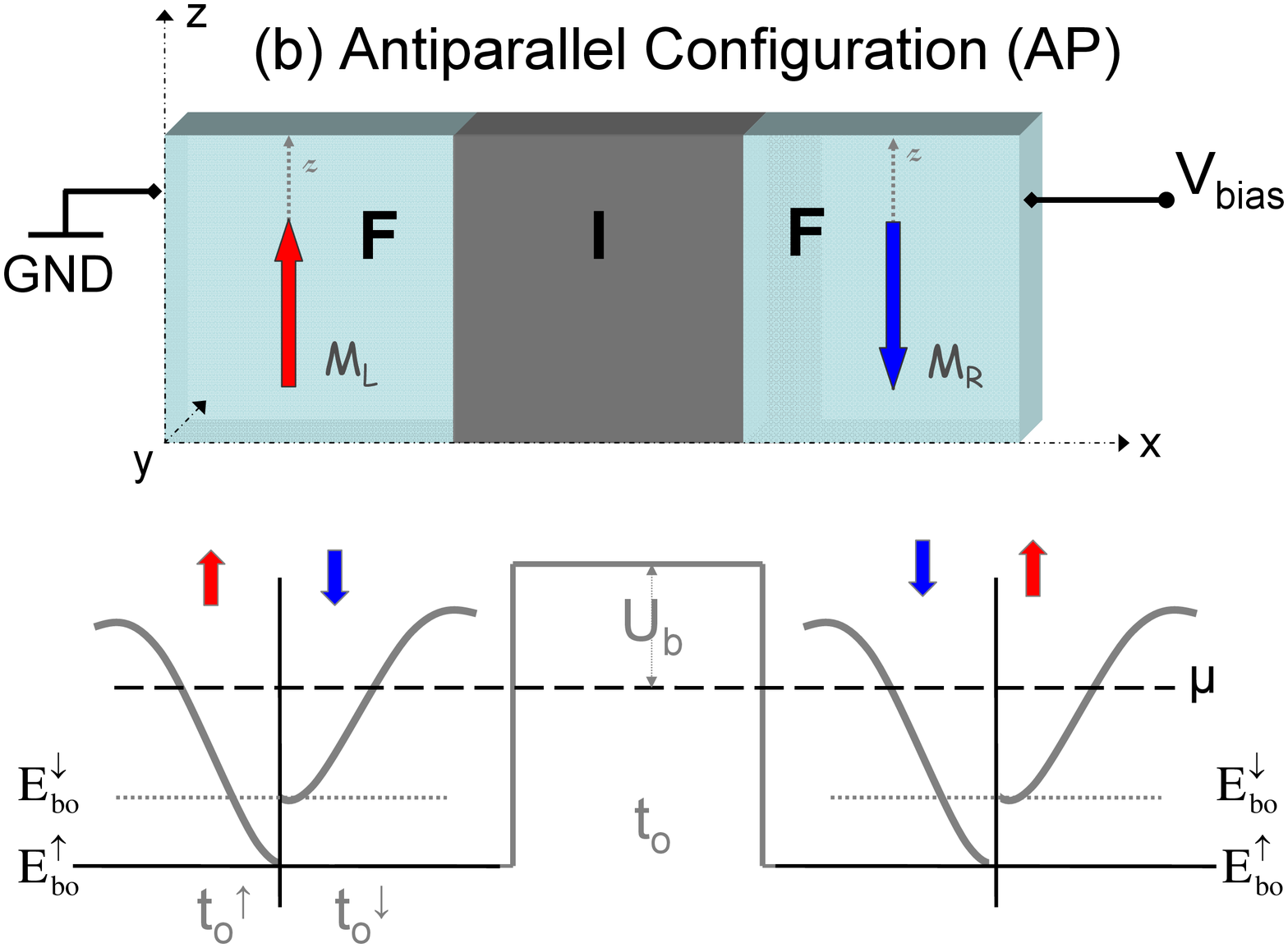}
\hskip -3.0in\includegraphics{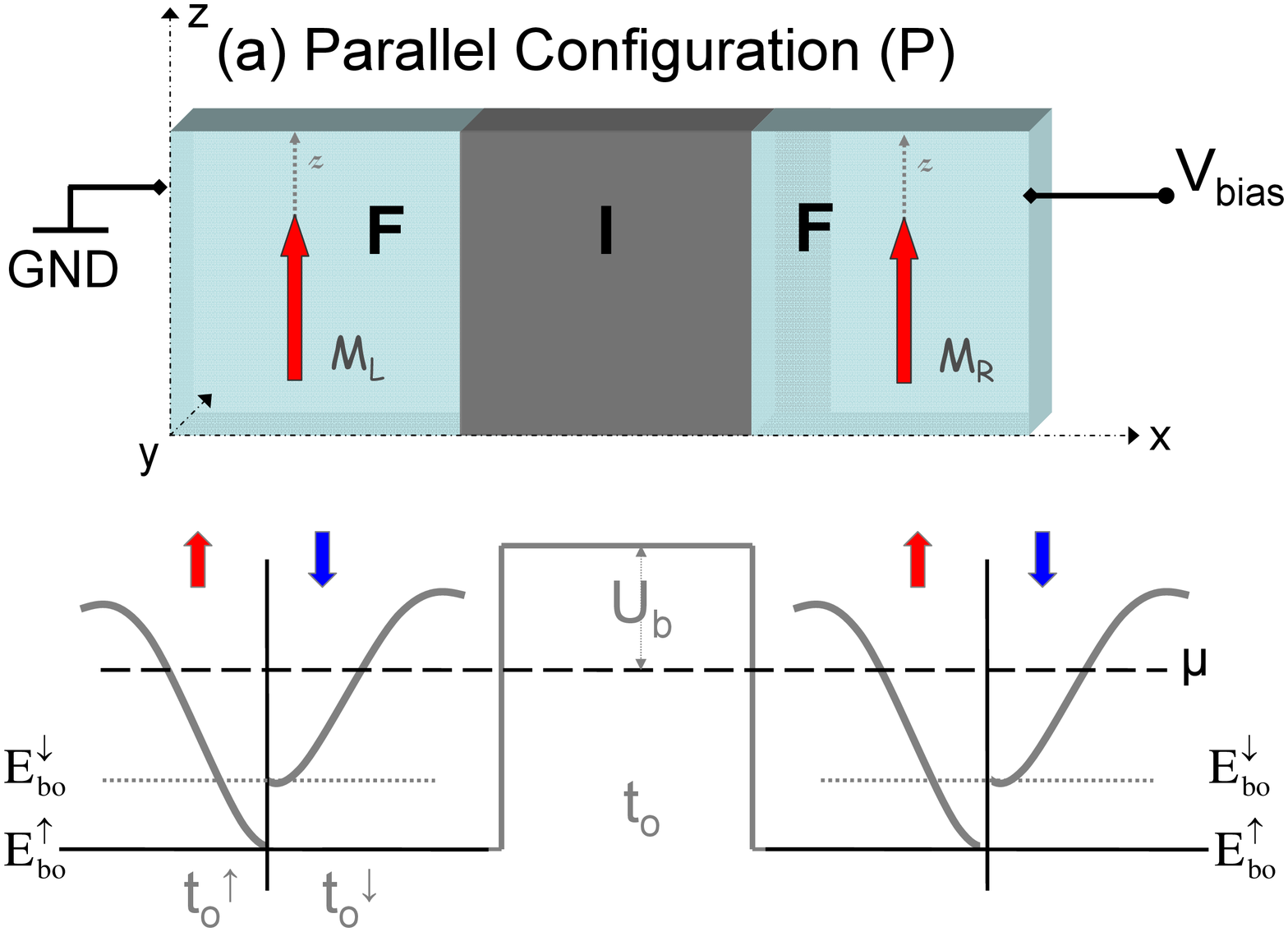}
\caption{A magnetic tunnel junction (MTJ) device. An insulating tunneling barrier is sandwiched between two metallic ferromagnetic layers. (a) Parallel (P) configuration along with its energy band diagram showing a cosine dispersion obtained using single-band tight-binding method. The magnetization of the two contacts are in the same direction. (b) Anti parallel (AP) configuration is shown, where the magnetizations of the two contacts are in opposite directions.}
\end{figure*}

\begin{figure}
\vspace{2.6in}
\hskip -3.8in\includegraphics{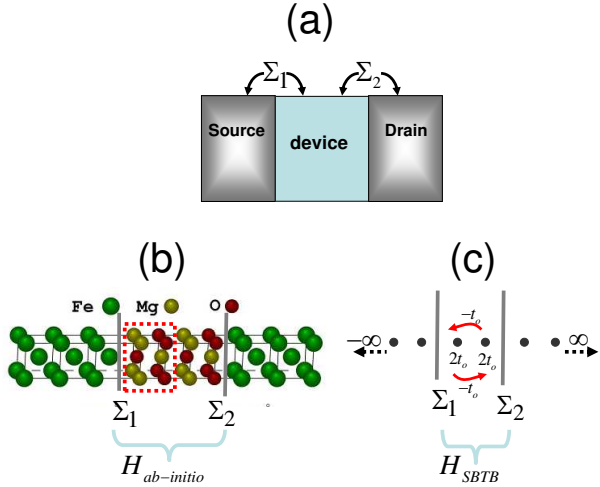}
\caption{The device structure for transport calculations. (a) A schematic of the device used. (b) The device structure used in \textit{ab-initio} calculations, reproduced from Ref. \cite{Heiliger05}. The unit cell is shown by dotted line which corresponds to one lattice point in SBTB-NEGF method. (c) Real space lattice for SBTB-NEGF calculations. First lattice points in Fe are treated as interface for which onsite Hamiltonian element is sum of hopping parameters of Fe and MgO and off diagonal elements are those of Fe or MgO depending on the neighboring lattice point. }
\end{figure}

In Ref. \cite{Raza07}, we report parameters for bcc Fe(100), which reflect the band dispersions for majority and minority spin bands over the two dimensional transverse Brillouin zone within a simple single-band tight-binding (SBTB) method \cite{Kittel_Book}. The transmission calculated using this method matches well with the one calculated using EHT-NEGF method (extended H\"uckel theory, non-equilibrium Green's function). In this paper, we propose parameters for MgO as shown in Table I and complement them with the Fe parameters in Ref. \cite{Raza07} to capture the band structure effects in Fe-MgO-Fe MTJ devices. This approach provides a simple and computationally inexpensive platform to explore MTJ devices. Furthermore, it gives an inherently simple and intuitive understanding for the underlying device physics.

The schematic of a MTJ device and the band diagram is shown in Fig. 1(a) and Fig. 1(b) for parallel (P) and anti-parallel (AP) configurations respectively. In P configuration, the magnetizations of the two contacts are in the same directions and in AP configuration, the magnetizations are opposite. Therefore, the P configuration has higher current densities than AP configuration. This change in current densities defines the TMR as $(J_P-J_{AP})/J_{AP}$. Higher TMR signifies a higher signal-to-noise ratio and hence is desirable. In this paper, we normalize TMR as $(J_P-J_{AP})/(J_{P}+ J_{AP})$ to compare our results with Refs. \cite{Heiliger05, Heiliger06} . 

For quantum transport, we use NEGF formalism \cite{Datta_Book}. For each band, we start with a one dimensional single-band tight-binding Hamiltonian \cite{Kittel_Book} for the device shown in Fig. 2:
\begin{eqnarray} H_{SBTB}=\begin{cases}E_{bo}+2t_o+U_L(i,j)\ for\ i=j\cr -t_o\ \ \ \ \ \ \ \ \ \ \ \ \ \ \ \ \ \ \ \ \ \ for\ |i-j|=1\end{cases}\end{eqnarray}
where $E_{bo}$ is the band offset and $U_L$ is the Laplace potential linearly dropped across the insulator region. For MgO, the hopping parameter $t_o$  is given in Table I and $t_o$ for Fe bands are given in Ref. \cite{Raza07}. Each device lattice point in Fig. 2(c) corresponds to a unit cell in Fig. 2(b) [shown by the dotted box]. The resulting dispersion is of the form $\epsilon(k)=E_{bo}+2t_{o}[1-\cos(ka_{l})]$, where $a_{l}=4.2\AA$ is the cubic lattice constant for MgO. At Fe-MgO interface, on-site elements are taken as average of $t_o$ for Fe and MgO. The off-diagonal elements are taken such that the resulting Hamiltonian is Hermitian as needed to ensure that the energy eigenvalues are real and current is conserved. The Green's function is then calculated as:
\begin{eqnarray}\hat{G}=[(E_l+i0^+)I-H_{SBTB}-\Sigma_c]^{-1}\end{eqnarray}
where $\hat{\Sigma_{c}}=-t_oe^{ik_{x,c}a_{l}}$ and c=[1,2] for the left and right contact respectively. The transmission is given as $\hat{T}(E_l)=tr(\hat{\Gamma_{1}}\hat{G}\hat{\Gamma_{2}}\hat{G^{\dagger}})$, which is $k_{||}$ independent and transmission per unit area is calculated as:
\begin{eqnarray}T_{SBTB}=\frac{1}{4\pi^2}\int_{(-\frac{\pi}{a},-\frac{\pi}{a})}^{(\frac{\pi}{a},\frac{\pi}{a})}dk_{y}dk_{z}\hat{T}(E_l)=\frac{1}{a^2}\hat{T}(E_l)\end{eqnarray} 
where $a=2.86\AA$ is the Fe cubic lattice constant. Current density in P and AP configuration for each spin orientation is then computed as:
\begin{eqnarray}J = \frac{e}{h}\int dE_l\ T_{SBTB}\ [f_{1}-f_{2}]\end{eqnarray}

\begin{figure}
\vspace{2.75in}
\hskip -3.8in\includegraphics{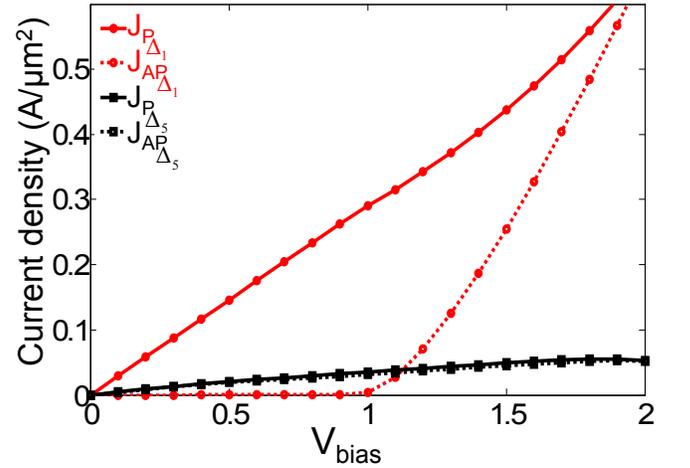}
\caption{Current density for $\Delta_{1}$ and $\Delta_{5}$ bands in Fe-MgO-Fe MTJ device. Current densities for parallel and anti-parallel configurations using SBTB-NEGF method. The parameters for MgO are shown in Table I and for Fe are in Ref. \cite{Raza07}.}
\end{figure}

\begin{table}
\caption{\label{tab:Table1} Parameters for MgO. $U_{b}$ is the MgO barrier height, $t_{o}$ is the hopping parameter. The parameters for Fe are in Ref. \cite{Raza07}.}
\begin{ruledtabular}
\begin{tabular}{lcc}
                              & $t_{o}(eV)$                 &  $U_{b}(eV)$         \\
\hline
$\Delta_{1}$ band             & 1.05                    &  2.6             \\
$\Delta_{5}$ band             & 1.05                    &  3.8             \\
\end{tabular}
\end{ruledtabular}
\end{table}

For $\Delta_1$ band, $U_b$ is taken to be half of the band gap in local density approximation (LDA) of the density functional theory \cite{Bredow00}. Although LDA underestimates band gap, we use this value to compare our results with Refs. \cite{Heiliger05, Heiliger06}. $U_b$ for $\Delta_5$ band is used as a fitting parameter. $t_o$ for both the bands are estimated by fitting current levels to Ref. \cite{Heiliger06}. The calculated current densities for $\Delta_1$ and $\Delta_5$ bands using these parameters are shown in Fig. 3.  

In Fig. 4, the total P ($J_{P-\textit{total}}$) and AP ($J_{AP-\textit{total}}$) current density is shown which match well with \textit{ab-initio} results \cite{Heiliger06}. $J_{P-\textit{total}}$ is dominated by $\Delta_1$ band current density. This is due to the lower potential barrier seen by the $\Delta_{1}$ band. On the other hand, the total AP current density is dominated by $\Delta_{5}$ band till about 1V. After this voltage, there is a sharp increase in AP current of $\Delta_1$ band and this band current density starts to dominate the total current density. This is due to half-metallic nature of $\Delta_1$ band. To elaborate more, any current in a particular configuration is made up of two components - the majority spin current $J_{\uparrow}$ and the minority spin current $J_{\downarrow}$. In anti-parallel configuration, both these current channels are made up of a minority band in one contact and a majority band in the other as shown in the band diagram in Fig. 1(b). At voltage below about 1V, current is small because the band edge of one of the contacts is still not in the bias window. Once the applied bias is high enough, the band is pulled within the bias window and the current starts to increase. In Fig. 5, TMR calculated using SBTB-NEGF method is shown. TMR values match well with those obtained using the \textit{ab-initio} model in Ref. \cite{Heiliger05}. The rapid decrease in TMR after about 1V is also consistent with other \textit{ab-initio} studies for a three layer device \cite{Derek07}. The TMR calculated using SBTB-NEGF method also follows a similar bias dependent trend as reported in experiments \cite{Tiusan06}. To the best of our knowledge, such a bias dependence and quantitative agreement has not yet been captured within a simple single-band tight-binding model or an effective mass based model. 

\begin{figure}
\vspace{2.75in}
\hskip -3.8in\includegraphics{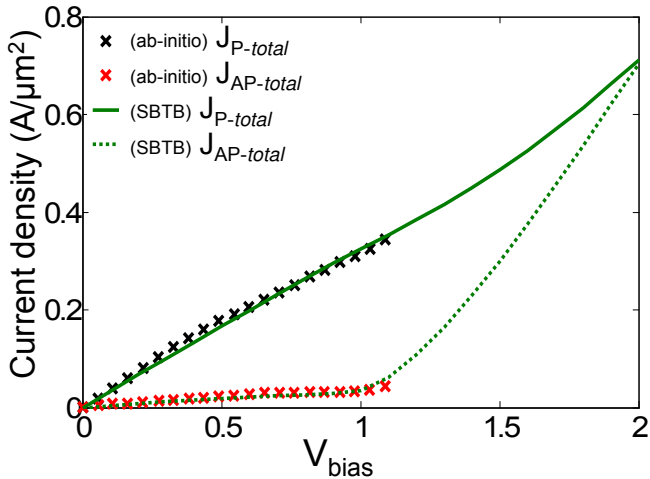}
\caption{Total current densities for P and AP configurations. The total current densities using SBTB-NEGF method for P and AP configurations are shown by solid and dotted lines respectively. The crosses are data from Ref. \cite{Heiliger06}. The low bias transport in P configuration is dominated by $\Delta_1$ band and in AP configuration is dominated by $\Delta_5$ band. The high bias transport in both configurations is dominated by $\Delta_1$ band.}
\end{figure}

\begin{figure}
\vspace{2.75in}
\hskip -3.8in\includegraphics{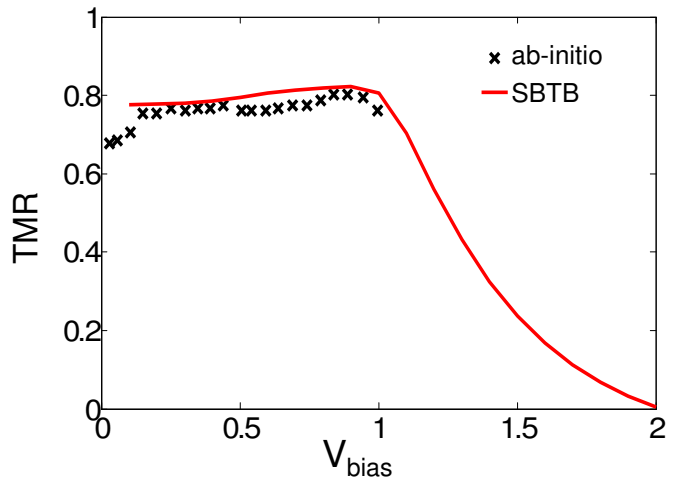}
\caption{Tunnel magnetoresistance (TMR). TMR calculated using SBTB-NEGF method is shown. x's show TMR calculated using \textit{ab-initio} method reproduced from Ref. \cite{Heiliger05}. The bias dependence of TMR is captured well within this simple single-band tight-binding model.}
\end{figure}

We have presented a single-band tight-binding model for bcc Fe-MgO-Fe magnetic tunnel junctions in [100] direction. We have tried to capture the band structure effects by using band parameters for different symmetry bands in the contacts and the barrier region. Features in TMR which are manifestation of the electronic structure of material were captured quantitatively within a simple model.

We thank Professor Supriyo Datta for very useful discussions. T. Raza acknowledges support by the MARCO Focus Center for Materials, Structure and Devices. H. Raza is thankful to National Science Foundation (NSF) and to Nanoelectronics Research Institute (NRI) through Center for Nanoscale Systems (CNS) at Cornell University. T. Raza is also thankful to E. C. Kan for providing office space and resources.


\begin{thebibliography}{100}
\bibitem{Butler01} W. H. Butler, X. -G. Zhang, T. C. Schulthess, and J. M. Maclaren, Phys. Rev. B {\bf{63}}, 054416 (2001).
\bibitem{Mathon01} J. Mathon, and A. Umerski, Phys. Rev. B {\bf{63}}, 220403R (2001).
\bibitem{Yuasa04} S. Yuasa, T. Nagahama, A. Fukushima, Y. Suzuki and K. Ando, Nature Materials {\bf{3}}, 868 - 871 (2004).
\bibitem{Parkin04} S. S. P. Parkin, C. Kaiser, A. Panchula, P. M. Rice, B. Hughes, M. Samant and S.-H. Yang, Nature Materials {\bf{3}}, 862 - 867 (2004).
\bibitem{Heiliger05} C. Heiliger, P. Zahn, B. Y. Yavorsky and I. Mertig, Phys. Rev. B {\bf{72}}, 180406(R) (2005).
\bibitem{Heiliger06} C. Heiliger, P. Zahn, B. Y. Yavorsky and I. Mertig, Phys. Rev. B {\bf{73}}, 214441 (2006).
\bibitem{Derek07} D.Waldron, L. Liu and H. Guo, Nanotechnology {\bf{18}}, 424026 (2007).
\bibitem{Raza07} T. Raza, H. Raza and S. Datta, In preparation. 
\bibitem{Kittel_Book} C. Kittel, \textit{Introduction to Solid State Physics} (Wiley, New York, ed. 7, 1996).
\bibitem{Datta_Book} S. Datta, \textit{Quantum Transport: Atom to Transistor} (Cambridge University Press, Cambridge, UK, 2005).
\bibitem{Bredow00}T.Bredow and A.R.Gerson, Phys.Rev.B {\bf{61}}, 5194 (2000).
\bibitem{Tiusan06} C. Tiusan, M. Sicot, M. Hehn, C. Belouard, S. Andrieu, F. Montaigne and A. Schuhl, Appl. Phys. Lett. {\bf{88}}, 062512 (2006).
\end{thebibliography}
\end{document}